\documentclass[%
  reprint,
superscriptaddress,
preprintnumbers,
 amsmath,amssymb,
 aps,
prb,twocolumn
]{revtex4-1}
\usepackage[normalem]{ulem}
\usepackage{amsmath}
\usepackage{amsfonts}
\usepackage{amssymb}
\usepackage{bm}
\usepackage{adjustbox}
\usepackage{epstopdf}
\usepackage{color}
\usepackage{xcolor}
\usepackage{textcomp}
\usepackage{gensymb}
\usepackage{subfigure}
\usepackage{graphicx}
\usepackage{dcolumn}
\usepackage{bm}
\usepackage{hyperref}
\hypersetup{colorlinks=true,allcolors=blue}
  \def \ie {{\it i.e.} }

\begin{document}
\title{Resistively detected NMR as a probe of the topological nature of conducting edge / surface states}

\author{Zekun Zhuang}
\affiliation{Department of Physics,   Brown University, Providence, Rhode Island 02912-1843, USA}
\author{V. F. Mitrovi{\'c}}
\affiliation{Department of Physics,   Brown University, Providence, Rhode Island 02912-1843, USA}
\author{J. B. Marston}
\affiliation{Department of Physics,   Brown University, Providence, Rhode Island 02912-1843, USA}
\affiliation{Brown Theoretical Physics Center, Brown University, Providence, Rhode Island 02912-1843, USA}

\date{\today }

\begin{abstract}
 
  Electron spins in edge or surface modes of topological insulators (TIs) with strong spin-orbit coupling cannot be directly manipulated with microwaves due to the locking of electron spin to its momentum. We show by contrast that a resistively detected nuclear magnetic resonance (RDNMR) based technique can be used to probe the helical nature of surface conducting states.  In such experiments, one applies a radio frequency (RF) field to reorient nuclear spins that then couple to electronic spins by the hyperfine interaction.  The spin of the boundary electrons can thereby be modulated, resulting in changes in conductance at nuclear resonance frequencies. Here, we demonstrate that the conductivity is sensitive to the direction of the applied magnetic field with respect to the helicity of the electrons.  This dependence of the RDNMR signal on angle probes the nature of the conductive edge or surface states. In the case of 3D TI in the quantum Hall regime, we establish that the dominant mechanism responsible for the conductance change in a RDNMR experiment is based on the Overhauser field effect. Our findings indicate that the same physics underlying the use of RDNMR to probe TI states also enables us to use RF control of nuclear spins to coherently manipulate topologically protected states which could be useful for a new generation of devices.

\end{abstract}

\maketitle


\section{Introduction}
The discovery of time-reversal-invariant topological insulators (TIs), which possess finite band gaps in the bulk and have gapless helical states on  boundaries, has attracted much attention in recent years \cite{Hasan2010,Qi2011,ando2013}. The novel features of TIs are intimately related to the underlying topology of the single-particle spectrum in momentum space, and are robust as long as the time-reversal symmetry is unbroken \cite{Kane20051,Kane20052,Fu2006,Fu20071,Fu20072}. The first two-dimensional TI was realized in HgTe/CdTe quantum well \cite{Bernevig2006,Konig2007}. The existence of topological edge state was confirmed by observing quantized longitudinal conductance. Later three dimensional TIs were also discovered in various materials, such as in $\text{Bi}_{1-x}\text{Sb}_x$ and $\text{Bi}_{2-x}\text{Sb}_{x}\text{Te}_{3-y}\text{Se}_y$ compounds \cite{Fu20071,hsieh2008,zhang2009,xia2009,Chen2009,Hsieh20091,hsieh20092,Brune2011,Jiang2012}. Angle-resolved photoemission spectroscopy (ARPES) and scanning tunneling microscopy/spectroscopy (STM/STS), provide direct evidence of the two-dimensional  (2D) Dirac fermions on the surface \cite{hsieh2008,xia2009,Chen2009,Hsieh20091,hsieh20092,Jiang2012}.

NMR techniques are a powerful probe of the electronic degrees of freedom due to the existence of hyperfine coupling. Spin polarized electrons create an effective local magnetic field that changes the nuclei resonance frequency, an effect that is easily detected. The technique has been applied to study the physics of quantum Hall effect, and electronic polarization in quantum wells of 2D electron gases and in quantum wires \cite{Dobers1988,Tycko1995,Barrett1995,Kuzma1998,Desrat2002,Mitrovic2007,Kou2010,Wald1994,Machida2002,Kronmuller1999,Stern2004,Gervais05,Kumada2007,Tiemann14,Khandelwal2001,Stern2012,Fauzi12,yang2017,Nisson2014}. 
NMR has also been used to investigate bulk properties of topological materials with non-local order \cite{Koumoulis2013,Mukhopadhyay2015,podorozhkin2015,Georgieva2016,tian2019,GUEHNE2019,papawassiliou2020}. The use of traditional NMR techniques to study properties of reduced dimensionality systems, such as GaAs heterostructures and surface states, is precluded as relatively few nuclear spins participate, making the signal of RF response weak. Various methods have been proposed to remedy the low sensitivity \cite{Dobers1988,Tycko1995,Barrett1995,Kuzma1998,Desrat2002,Mitrovic2007}.  Resistively Detected NMR (RDNMR) has been the most powerful tool developed thus far to probe the fundamental physics of the reduced dimensionality systems.  In  confined geometries the overlap of the electronic degrees of freedom with the nuclei can be relatively large resulting in strong hyperfine interaction between electronic and nuclear spins.  Strong  hyperfine coupling in quantum wells was exploited in RDNMR to probe the quantum Hall-regime \cite{Dobers1988,Kronmuller1999,Desrat2002,Stern2004,Gervais05,Kumada2007,Tiemann14,Tracy2006,Tracy2007}. In quantum Hall systems, the nuclei are first polarized statically by application of a strong magnetic field, or dynamically, for instance through the spin flip-flop processes at quantum point contacts \cite{GlazovHFBook}. The nuclei are then depolarized by irradiation with a resonant RF field, which can then be seen in a variation of the resistance that may be attributed to either nucleus-electron scattering or a change of the Overhauser field \cite{Overhauser1953}. Sensitive RDNMR measurements can thus uncover rich structures of electronic states and the interplay between electrons and nuclei.

In this paper, we propose RDNMR experiments that can be performed in TI systems as a new tool to identify and characterize the Dirac fermions on the edges or surfaces. An advantage of RDNMR, compared to the ARPES and STM/STS experiments, is that it also provides a way to directly control the Dirac fermions. The local nature of hyperfine interaction permits the direct manipulation of the electronic spin, even when the electronic spin is coupled to its orbital motion, as is the case in the TIs. 
This is particularly useful because electronic spin is not a good quantum number when there is strong spin-orbit coupling, and resonant microwaves cannot control the electron spin polarization.
Here we identify  two different mechanisms giving rise to RDNMR in TIs. These are spin flip-flop processes and the Overhauser field effect. For the former type, because of the helical nature of the edge states, nuclei on the edge are dynamically polarized along the quantization axis determined by a weak external magnetic field when bias is applied \cite{Lunde2012,Probst2015,Russo2018}. When subjected to a RF field, the depolarized nuclei backscatter the Dirac fermions on the edge/surface and change the current. The second mechanism can be applied to a 3-D TI under a strong magnetic field, that is, in the quantum Hall regime, which is analogous to the mechanism found in \mbox{Ref. \onlinecite{Desrat2002}}: the effective magnetic field produced by nuclei, the Overhauser field, varies when nuclei are in resonance and this modifies the gap between adjacent Landau levels, which leads to a change of surface carrier densities and hence the transport properties.

The paper is organized as follows. In Sec. \ref{Sec2DTI}, we introduce the dynamic nuclear polarization on the edges of a 2-D TI, discuss how RDNMR works through spin flip-flop processes and compare the experimental signals that could distinguish helical edge states and non-helical edge states. We then generalize this idea to a clean 3-D TI wire in Sec. \ref{Sec3DTI} and demonstrate similar results. In Sec. \ref{Sec3DTI}, we also discuss how RDNMR based on Overhauser shift effect can be realized in 3-D TIs in the quantum Hall regime. Discussions and conclusions are presented in Sec. \ref{SecDis}. 

\section{Two dimensional topological insulators} \label{Sec2DTI}
In this section, our calculations regarding the presence of the RDNMR response in 2D TIs is described. We find that RDNMR response exists and exhibits a dependence on the orientation of the applied magnetic field only if the edge states are helical.

In 2-D TIs, there are fermions with opposite spins counterpropagating on the edges of the system, which are called helical edge states. Although the details of the edges and microscopic parameters vary from systems to systems, the helical edge state can be well described by a minimal Hamiltonian at low energies:
\begin{equation}
H_{\text{edge}}=\hbar v_F k \sigma_z, \label{EdgeH}
\end{equation}
where $\sigma_z$ is the Pauli spin matrix and $v_F$ is the Fermi velocity. We assume that Eqs. (\ref{EdgeH}) and (\ref{3DH}) (see below) can still serve as minimal Hamiltonians that capture the essential physics even in the presence of a weak external magnetic field or the Overhauser field due to polarized nuclei.

We now discuss the meaning of the `weak' and `strong' magnetic fields terminology that is used in the paper. A magnetic field is weak if it does not drastically change the low energy physics on the edges or surfaces of a TI. The field only determines the quantization axis of nuclei and is used to perform NMR experiments. For 2-D TIs, we assume that the edge physics is still well described by helical edge states upon the application of such weak field. We do not consider, for example, edge excitations that become gapped or helical edge states that transition to chiral edge states at the Fermi level \cite{Tkachov2010,Chen2012,Scharf2012}. For 3-D TIs considered later, the magnetic field may be  considered to be weak if the temperature is much larger than the Landau gap or Zeeman gap so the surface states may still be approximately described by gapless 2-D Dirac fermions. In contrast, a strong magnetic field not only polarizes nuclei but also drastically changes the low energy physics of a TI. In our work this is the case for 3-D TIs in the quantum Hall regime, where the conducting surfaces that is perpendicular to the magnetic field become gapped due to the formation of Landau levels.

The interaction between helical fermions and nuclei on the edge is modelled by the Fermi-contact-type hyperfine interaction,
\begin{equation}
H_{\text{hf}}=\sum_{i,s}\left( A_z^{s} I^s_z S_z+ A_{{xy}}^{s}\frac{I^s_- S_++I^s_+ S_-}{2} \right)\delta^{(D)}(\vec{r}- \vec{R}_i), \label{HfH}
\end{equation}
where $\vec{R}_i$ is the coordinate of nuclei, $s$ labels the isotope species, $A_{z({xy})}^{s}$ is the anisotropic out(in)-plane hyperfine coupling constant for isotope $s$ , $I^s_{z,+,-}$ are the dimensionless nuclear spin operators and $S_{z,+,-}$ are the dimensionless electronic spin operators. For a 2-D TI, the nuclei interact with electrons only on the 1-D edge so $D = 1$, while for a 3-D TI $D$ equals 2 since the Dirac fermions live on the 2-D surface. Note that although in this work we assume the electronic density to be constant on the edge/surface for convenience, any variation of it can be absorbed into the $A_{z({xy})}^{s}$ and hence does not affect the final results.

\begin{figure} 

    \centering
    \includegraphics[width=0.50 \textwidth]{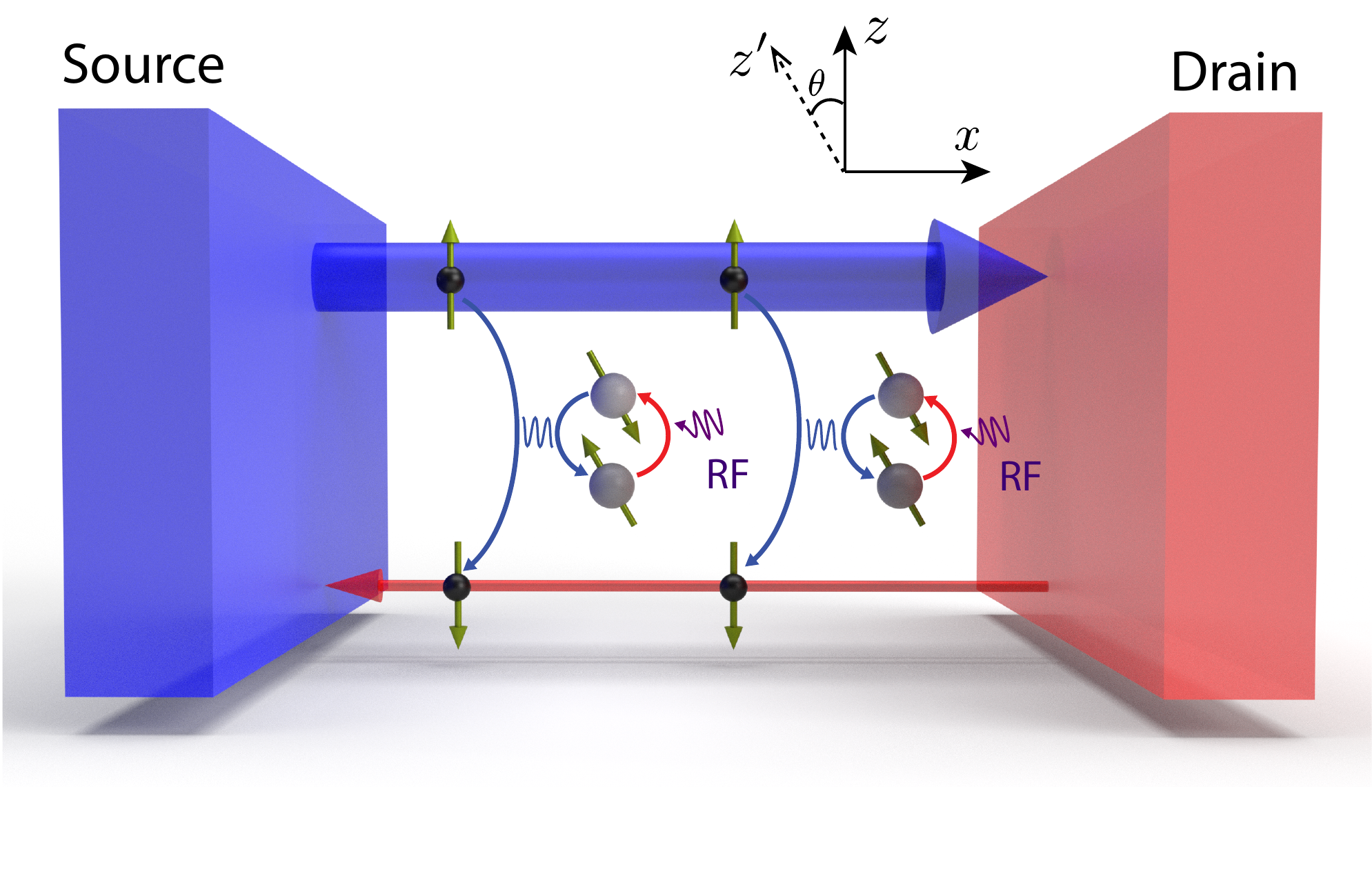}
    \vspace*{-1.2cm}
    \caption{Schematic of RDNMR for a 2-D TI. When a RF field is applied to depolarize/flip the nuclei, the nuclei continuously back-scatter the helical Dirac fermions through the hyperfine interaction. The static weak magnetic field is applied along the $\hat{z}^\prime$ direction.}
    \label{2Dplot}

\end{figure}

When the edge is short and a voltage bias is imposed there will be unequally populated right (spin-up) and left (spin-down) movers. The imbalance would tend to equilibrate through spin flip processes by the hyperfine interaction, as illustrated in Fig. \ref{2Dplot} by the blue interaction lines.
\begin{figure*}[!ht]
\centering
  \subfigure[]{
    \label{G:1} 
    \includegraphics[height=2.7 in]{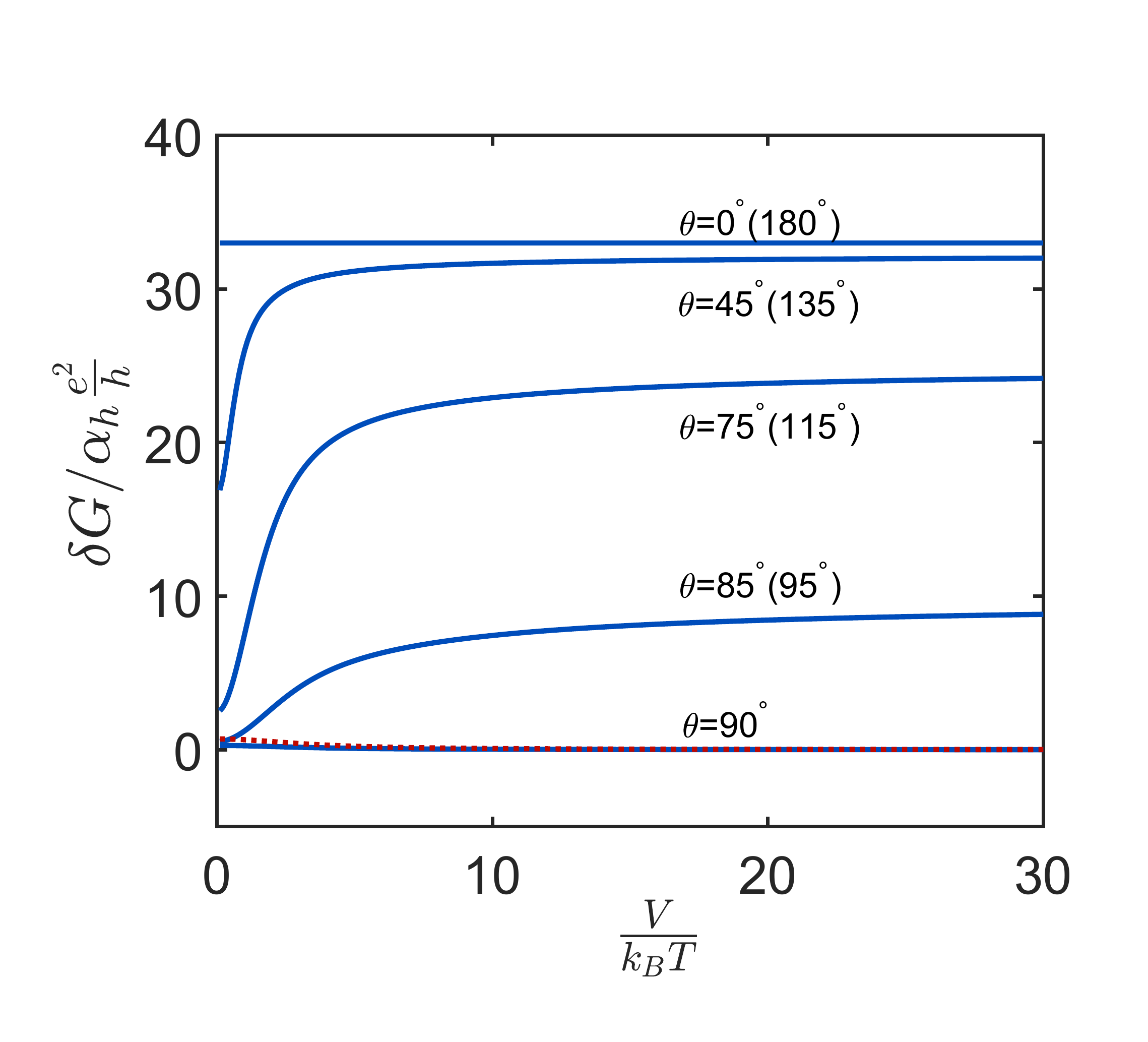}}
  \subfigure[]{
    \label{G:2} 
    \includegraphics[height=2.7 in]{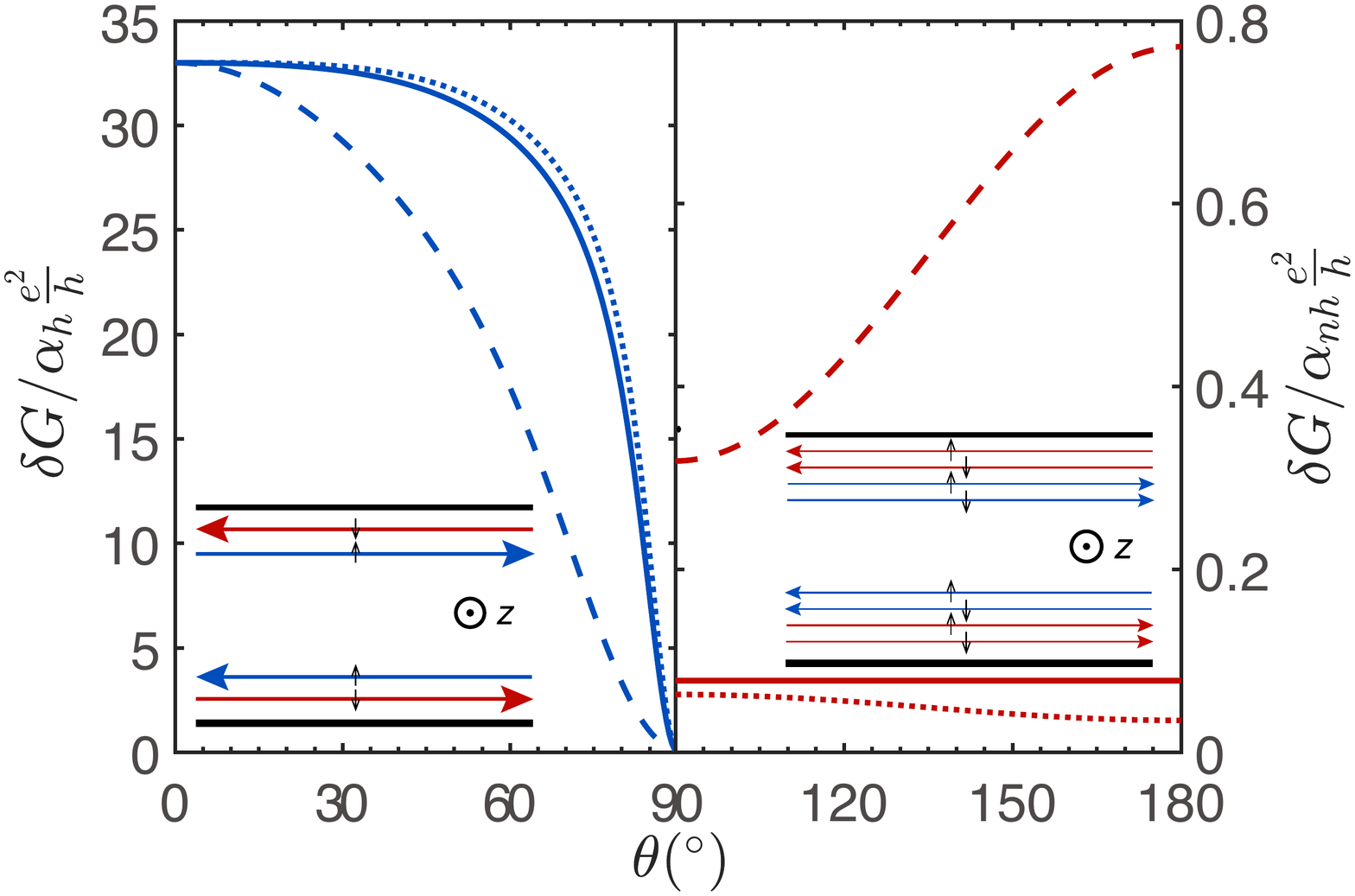}}

\caption{(a) The change of conductance at the NMR resonance frequency vs. $V/k_BT$ at different angles $\theta$ in a  two-terminal experiment, for isotropic hyperfine coupling $|A_z|=|A_{xy}|$. (b) The change of conductance at the NMR resonance frequency versus angle $\theta$ for different anisotropic hyperfine coupling $|A_z/A_{xy}|$. The solid line represents $|A_z/A_{xy}|=1$, the dashed line represents $|A_z/A_{xy}|=4$ and the dot line is for $|A_z/A_{xy}|=0.25$. The blue lines show the 2-D TI while the red lines represent the case when there is one spin-degenerate  conducting channel on the edge, as depicted in the insets of (b). In the insets, the black lines represent the edges of the sample while blue and red lines show the counter-propagating conducting channels. $\alpha_h=\pi^2 N_n |A_{xy}|^2/h^2v_F^2$, $\alpha_{nh}=\pi^2 N_n J^2 |A_{xy}|^2/16$. Parameters: $I=9/2$, $\Delta/k_BT=0.05$.}
\label{G}
\end{figure*}
Due to the helicity of the edge state and the conservation of total angular momentum, such process would gradually polarize the nuclear spins at the edges, which is referred as Dynamic Nuclear Polarization (DNP), until the whole system reaches non-equilibrium steady state (NESS) \cite{Lunde2012,Probst2015,Russo2018}. For the case that the nuclear spins relax only through hyperfine interaction with Dirac fermions at the edges, which we assume throughout this work, it may be shown that for spin-$1/2$ nuclei in the absence of external magnetic field, its magnetization is $M=\frac{1}{2}\tanh (\frac{V}{2k_B T})$, where $V=\mu_L-\mu_R$ and $\mu_{L(R)}$ is the chemical potential of left(right) reservoir  \citep{Lunde2012}. On the other hand, although there would be finite backscattering current during the transient period due to the electron-nucleus backscattering process, in the NESS there would be no backscattering current because the spin-flop rate of nuclear spins, which is also the scattering rate of right or left movers, vanishes in the absence of other nuclear relaxation mechanisms.

However, such a backscattering current can be maintained by performing a NMR experiment. 
(Throughout this paper, the phrase ``performing a NMR experiment'' implies that the polarized nuclear spin systems is subject to the resonant RF field.) That is, suppose that the static magnetic field  $\vec{B_s}=B\hat{z}$ is sufficiently weak not to affect the effective electronic Hamiltonians in Eqs. \ref{EdgeH} and \ref{3DH}, and a weak radio-frequency (RF) magnetic field perpendicular to $\vec{B_s}$ at the resonance frequency of isotope $s$ are applied, the nuclei of kind $s$ will be depolarized and may be regarded to have infinite spin temperature at NESS, if one assumes that the driving is much stronger than the relaxation for nuclei and nuclei have saturated. In this case, because these nuclei are no longer polarized and are allowed to scatter with the helical fermions, the helical fermions at edges would be continuously backscattered by those nuclei at resonance, as indicated in Fig. \ref{2Dplot}. The backscattering current $I_{sb}$ on a single edge in this case can be found using Fermi's golden rule (see Appendix \ref{App1}), that is

\begin{equation}
I_{bs}\approx  \alpha_h^s\frac{e}{h}\overline{(I^s-m^s)(I^s+m^s+1)}V,
\end{equation}
which leads to suppression of the conductance
\begin{equation}
\delta G\approx -\alpha_h^s\frac{2e^2}{3h}I^s(I^s+1), \label{2Dchange}
\end{equation}
compared to the ideal quantized conductance $G_0=\frac{2e^2}{h}$ in a two-terminal experiment, if $k_BT\ll V$ is assumed and the Zeeman splitting of nuclei is neglected. We use $\overline{(...)}$ to denote the average value with respect to specific population of isotope $s$, $\alpha_h^s=\frac{\pi^2 N_n^s |A_{xy}^s|^2}{h^2v_F^2}$, and $N_n^s$ is the number of specific isotope $s$ covered by edge state on the single edge. $I^s$ and $m^s$ are the spin quantum number and magnetic quantum number of nuclear species $s$ in the direction of magnetic field respectively. 

In general, the weak static magnetic field $\vec{B_s}=B\hat{z}^\prime$ can be applied along arbitrary direction, and in this case the nuclear spins would not always be fully polarized by the current \cite{Probst2015}. The population ratio of nuclei can be found to be $P_{m+1}^h/P_m^h=\cot^4(\theta/2)$ using Fermi's golden rule (see Appendix \ref{App1}), where $\theta$ is the angle between $\hat{z}$ and $\hat{z}^\prime$ (see Fig. \ref{2Dplot}), if again $k_BT\ll V$ is assumed and the Zeeman splitting of nuclei is neglected. One could therefore conclude that at angle $\theta=\pi/2$ ideally no change of conductance would be expected when RF field at resonance frequencies is applied, since the configuration of nuclear spins does not change before and after the application of RF field. Therefore, one would infer that the change of conductance when RF field is applied is anisotropic with respect to $\theta$, even when the hyperfine coupling constant is isotropic, i.e. $A_{xy}^s=A_z^s$. A full calculation with parameters $\theta$, $|A_z/A_{xy}|$ and $V/k_B T$ (see Appendix \ref{App1}), which also takes into account the effect of nuclear Zeeman energy $\Delta^s$, shows that for the two-terminal setup where the conductance is provided by both edges, the conductance change is in fact strongly anisotropic as displayed in \mbox{Fig. \ref{G}}. 

 To compare, we also consider the case in which non-helical fermions live at the edge of a 2-D trivial insulator, where each edge has one spin-degenerate conducting channel and bulk is insulating, as depicted in the inset of the right panel of Fig. \ref{G:2}. 
As evident in Fig. \ref{G}, if the edge state is non-helical ideally there would be no angular dependence in the RDNMR signal if the hyperfine coupling is isotropic, which is due to the spin degeneracy of particles at the edges. Furthermore due to the absence of effective DNP the conductance change of a non-helical fermionic system when NMR is turned on is much smaller than that of a helical system. Even though our result assumes that nuclei are depolarized by continuous-wave NMR, the results should be qualitatively the same if nuclei are flipped by pulse techniques.

\section{Three dimensional topological insulators} \label{Sec3DTI}
\begin{figure*}[!ht]
\centering\hypertarget{3Dsch}{}
    \includegraphics[width=0.8 \textwidth]{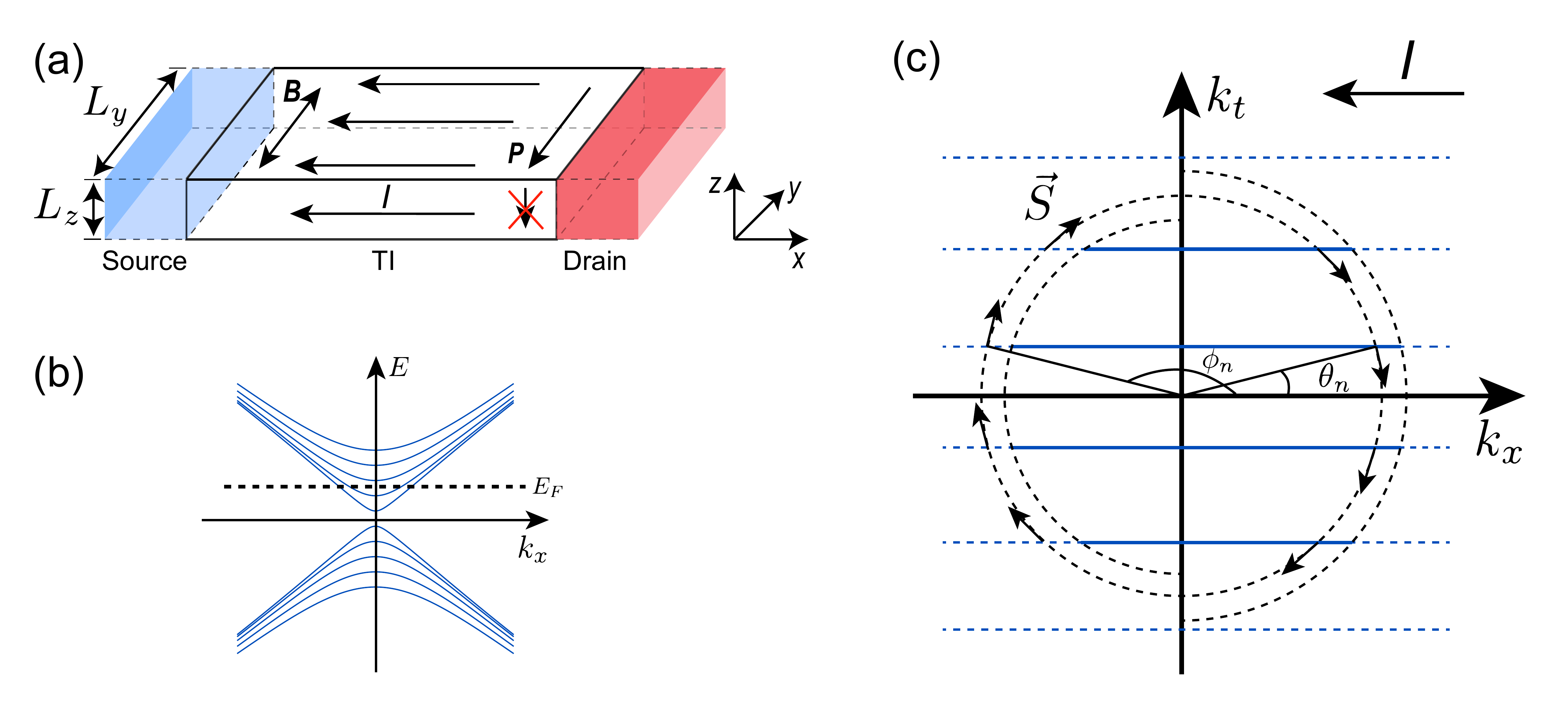}
\caption{(a) Schematic of the setup in our model. The arrows denote the directions of the magnetic field $B$, current $I$ and the nuclear polarization $P$ respectively. The double arrows on the magnetic field $B$ indicate that it can be aligned along either direction. In this case $P$ is non-vanishing only on top and bottom surfaces. (b) The surface spectrum of a rectangular wire under the quantization condition. (c) The schematic of the distribution of Dirac fermion when biased, where solid blue lines denote the occupied states while dashed blue lines represent the unoccupied ones; the full circle is the Fermi surface at equilibrium while the two hemi-circles denote the Fermi surface of the reservoirs when bias is present. }
\label{3Dsch}
    
\end{figure*}

It is an interesting question whether similar physics could also exist in a 3-D strong TI that hosts a single Dirac cone on the surface. This question is addressed in the calculations below. Our main finding indicates that RDMNR response in 3D TIs is present but may be weaker than in the 2D case as the current-induced DNP is  feeble.  Furthermore, in a 3-D TI that is in the quantum Hall regime, we demonstrate that the effect of the Overhauser field is to modify the Landau levels spectra and thereby induce an RDNMR response.
 
 We consider a 3-D rectangular TI wire, as shown in Fig. \ref{3Dsch}\hyperlink{3Dsch}{(a)}, with simplified surface Hamiltonian on all the surfaces
\begin{equation}
H_\text{surf}=\hbar v_F(\vec{\sigma}\times \vec{k})\cdot\hat{n}, \label{3DH}
\end{equation}
where $\hat{n}$ is the unit normal vector of the corresponding surface and we have neglected the hybridization between opposite surface states across the bulk. Note that Eq. (\ref{3DH}) is a highly simplified surface Hamiltonian which may give a very different spectrum from the real one in a nanowire \cite{Zhang2014}. This is not only because higher order terms are absent, but also due to the fact that in reality surfaces with different orientations typically have different Hamiltonians attributed to the bulk anisotropy of commonly used TI materials \cite{Zhang2012,Brey2014,Silvestrov2012}. Furthermore, the pseudo-spin $\vec{\sigma}$ in Eq. (\ref{3DH}) may be very different from the real spin $\vec{S}$ on some surfaces, although we simply assume $\vec{S}=\vec{\sigma}/2$ throughout this work \cite{Zhang2012}. Nevertheless, the Hamiltonian Eq. (\ref{3DH}) is the simplest model that captures the key physics and allows us to make explicit analytic calculations.  In such ideal model the surface state may be labelled by longitudinal momentum $k_x$ and the transverse momentum $k_t$ along the $y$-$z$ direction, which is quantized as $k_t=\frac{(2n+1)\pi}{2(L_y+L_z)}, n\in \mathbb{Z}$ \cite{Egger2010,Yar2019}. Therefore series of subband are developed, as depicted in Fig. \ref{3Dsch}\hyperlink{3Dsch}{(b)}. Note that this quantization condition is attributed to the accumulated $\pi$ Berry phase when a Dirac fermion travels around the cross-section. 

We consider the coherent transport along $x$ direction in a clean sample. For the two-terminal setup shown in Fig. \ref{3Dsch}\hyperlink{3Dsch}{(a)}, if there is no reflection at the contacts, the conductance is $G=Me^2/h$ according to the Landauer formula, where $M$ is the number of transverse modes. Comparable to the 2-D case, due to the unequally populated right and left movers, the nuclei would be polarized in a similar manner when bias is applied. In fact, take the top surface shown in Fig. \ref{3Dsch}\hyperlink{3Dsch}{(a)} as an example, the nuclei on the top surface would not be polarized at all by the current if the external magnetic field is along the $x$ or $z$ axis, but could be partially polarized if the external field is along the $y$ axis. Such anisotropy is due to the helicity of the Hamiltonian Eq. (\ref{3DH}): the spin is always in-plane and perpendicular to the momentum. When the magnetic field is along $y$ direction, the current-induced population imbalance prefers to flip the nuclei from $+\hat{y}$ direction to $-\hat{y}$ direction since the majority of right(left)-movers have considerable spin component along the $-\hat{y}$($+\hat{y}$) direction (see Fig. \ref{3Dsch}\hyperlink{3Dsch}{(c)}). It should be emphasized again that here the external magnetic field affects neither the electronic spectrum nor the nuclear polarization directly, instead it only determines the quantization axis of the nuclear spin. With detailed calculation based on Fermi's golden rule (see Appendix \ref{App2}) and simplification $V\gg k_BT\gg \Delta$, $A^s\equiv A_z^s=A_{xy}^s $, it can be shown that the population ratio of nuclei in this case is given by
\begin{equation}
    \frac{P_{m+1}}{P_{m}}=\left(\frac{\gamma}{1-\gamma}\right)^{2\cdot\text{sgn}\left(v_FV\right)} \label{DNP3D},
\end{equation}
where $m$ labels the eigenstate of $I_y$, \mbox{$\gamma=\sum J_n\sin^2 (\theta_n/2)/\sum J_n$} depends on the Fermi energy and the size of the sample, \mbox{$J_n= [hv_F(L_y+L_z)\cos\theta_n]^{-1}$} is the DOS of subband $n$ per surface area at Fermi level, and $\theta_n$ is the angle between $\vec{k}$ and $k_x$ axis that ranges between $-\pi/2$ and $\pi/2$ (see Fig. \ref{3Dsch}\hyperlink{3Dsch}{(c)}). In the case when $M\gg 1$, one could transform sum to integral and obtain $\gamma=1/2-1/\pi$ so that $P_m/P_{m-1}\approx 0.05$. This suggests that considerable nuclear polarization on the top surface may be achieved if the magnetic field is in-plane and normal to the electric field. One could deduce that in this case opposite nuclear polarization would appear on the bottom surface and no current-induced nuclear polarization would exist on the side surfaces, as shown in Fig. \ref{3Dsch}\hyperlink{3Dsch}{(a)}.

The possibility of having considerable DNP indicates that relatively large conductance change may be observable in 3-D strong TIs when NMR is performed. We continue assuming the external field is applied along $y$ direction. Similar to the 2-D case, the backscattering current due to the nuclei can be calculated in a similar manner and in fact we are only interested in the scattering due to the nuclei on the top and bottom surfaces, because only those nuclei could change appreciably when NMR is performed. The suppression of conductance due to the scattering with specific isotope $s$ on the top and bottom surfaces can be calculated to be
\begin{multline}
    \delta G=-\frac{\pi^2 N_n^s (A^s)^2 J_{\text{surf}}^2}{4}\frac{e^2}{h}\times \left[2\overline{(I_m^z)^2}\gamma(1-\gamma)\right.\\\left.+\overline{(I_m^+)^2}\gamma^2  +\overline{(I_m^-)^2}(1-\gamma)^2 \right], \label{deltaG3D}
\end{multline}
where $J_{\text{surf}}=\sum_n J_n$ is the DOS of surface state per surface area at Fermi level, $N_n^s$ is the number of specific isotope $s$ on the top and bottom surfaces and for simplicity we have assumed isotropic hyperfine coupling. We have defined $I^\pm_{m}=\langle I,m\pm 1\vert I_\pm^s\vert I,m\rangle=\sqrt{(I\mp m)(I\pm m+1)}$ and $I^z_m=m-\overline{m}$. The first term in the above equation is due to the nuclear-spin-conserving scattering while the last two terms correspond to the nuclear-spin-flipping scattering. Similar to the 2-D case, the change of conductance when the nuclei is at resonance can be calculated by computing the difference of conductance when nuclei are polarized and depolarized using Eq. (\ref{DNP3D}) and Eq. (\ref{deltaG3D}), which is approximately
\begin{equation}
 \delta G\approx-\frac{\pi^2 N_n^s (A^s)^2 J_{\text{surf}}^2}{6}I^s(I^s+1)\frac{e^2}{h} ,
\end{equation}
in the limit $\gamma \rightarrow 0$, \ie the nuclear spins are fully polarized by the current.

\begin{figure}[b]
\includegraphics[width=3in]{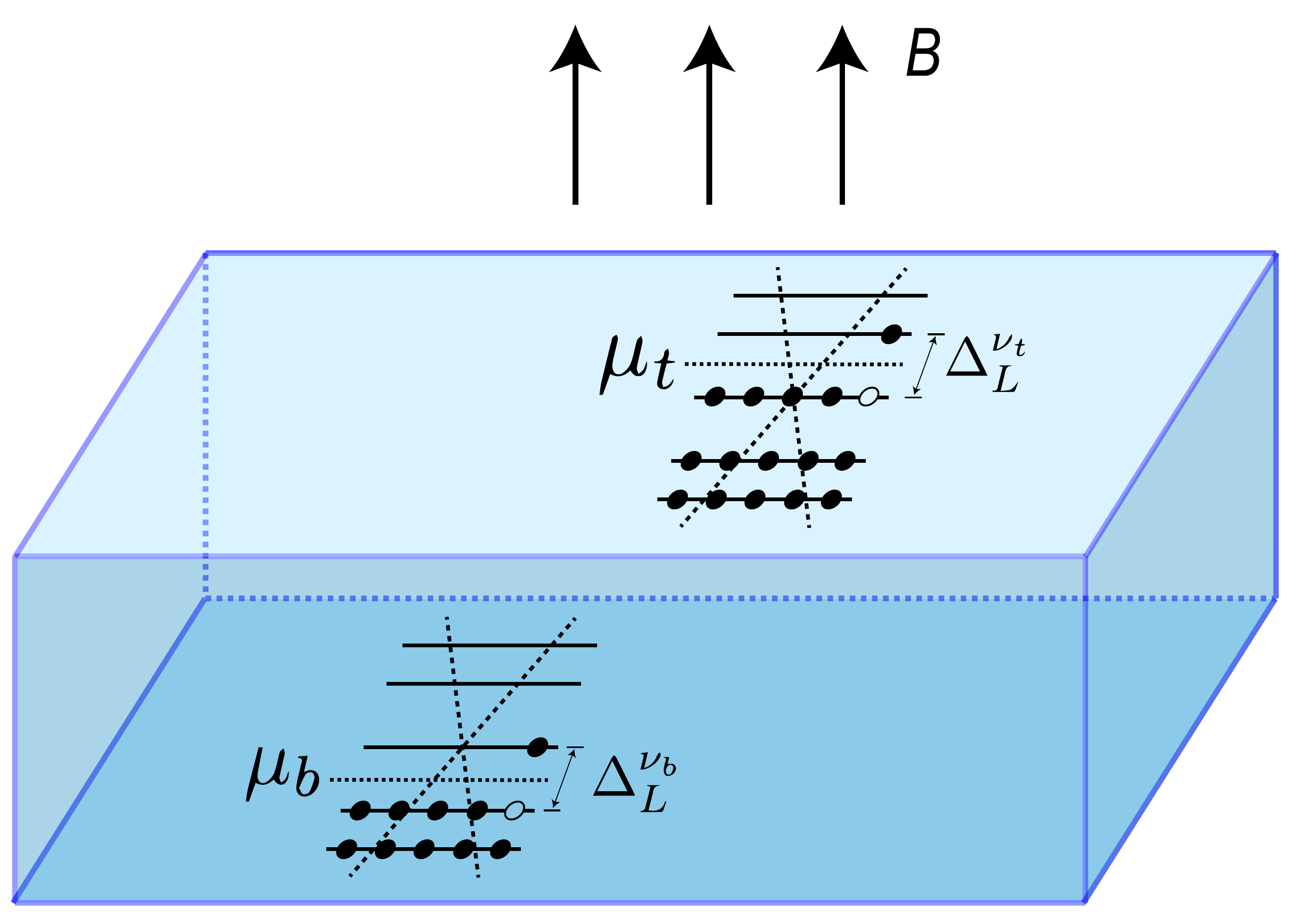} 
\caption{Schematic for a 3-D TI in a strong magnetic field. Landau levels are formed on the top and bottom surface with filling factor ($\frac{1}{2},-\frac{1}{2}$) as an example. Due to the thermal activation, there is nonzero carrier density contributing to conductivity on the top and bottom surface.}\label{3DTI}
\end{figure}

We propose an alternative mechanism, based on the Overhauser field effect, to control the conductivity by performing NMR in 3-D strong TIs in the presence of strong magnetic field. In this case, the magnetic field not only determines the quantization axis of nuclear spin but also modifies the Hamiltonian Eq. (\ref{3DH}) drastically \cite{Liu2010,Lee2009,Vafek2011,Xie2011,Brey2014,zhang2015}. Under an external magnetic field $B_z$ along $z$ direction, as depicted in Fig. \ref{3DTI},
 due to both the orbital effect and Zeeman effect of the magnetic field, the spectrum of Dirac fermion on the top and bottom surface quantizes to well-known Landau levels \citep{Liu2010}:
\begin{equation}
E_{\pm N}=\left\lbrace \begin{array}{ll}
\pm\sqrt{(g_z\mu_B B_\text{eff}/2)^2+2\hbar v_F^2Ne\vert B_z\vert},& N\geqslant 1\\
-g_z\mu_B \vert B_\text{eff}\vert/2,& N=0\\
\end{array}\right. , \label{LL}
\end{equation}
while the spectrum on the side surface only shifts by a finite amount in momentum space. In Eq. (\ref{LL}) \mbox{$B_\text{eff}=B_z+B_n$} incorporates both the external magnetic field $B_z$ and the macroscopic Overhauser field $B_n$ produced by nuclei, $g_z$ is the effective $g$-factor of electron along $z$ direction and $\mu_B$ is the Bohr magneton. We note that since Overhauser field does not contribute to the orbital effect and only contributes to the Zeeman term in Eq. (\ref{LL}), the change of it does not affect the filling factor. Due to the presence of such quantized Landau levels, in a 3-D TI thin film subject to high magnetic field and at low temperature, integer quantum Hall effect with quantized Hall conductance $\sigma_{{xy}}=(\nu_t+\nu_b)e^2/h$ appears at half-integer top and bottom filling factor, $\nu_t$ and $\nu_b$, which may be tuned independently by gate technique \citep{xu2014,yoshimi2015,xu2016,Chong2019}. In traditional quantum Hall systems, it is well known that in the thermal activation regime the longitudinal conductivity obeys Arrhenius-type law, namely $\sigma_{xx}\propto e^{-\Delta_L/(2k_B T)}$ \cite{Polyakov1994}. It is therefore expected that, for a 3-D TI in the quantum Hall regime, if the cyclotron gap on the top and bottom surfaces change by a small amount $\delta \Delta_L^{\nu_t}$ and $\delta \Delta_L^{\nu_b}$, the change of longitudinal conductivity would be
\begin{equation}
\delta \sigma_{xx}\approx \sigma_t^0(e^{-\delta \Delta_L^{\nu_t}/2k_BT}-1)+\sigma_b^0(e^{-\delta \Delta_L^{\nu_b}/2k_BT}-1), \label{deltaSigma}
\end{equation}
where $\sigma_{t(b)}^0$ is the conductivity contributed by top(bottom) surface before the change of cyclotron gap. Such a cyclotron gap can be controlled by tuning the Overhauser field through NMR techniques. At thermal equilibrium, nuclei have polarization due to the strong magnetic field. When specific isotope $s$ is completely depolarized by RF field, the Overhauser field $B_n$ changes by \cite{abragam1961}
\begin{equation}
\delta B_n\approx -\frac{I^s(I^s+1)}{3}\frac{n^s \gamma_n^sA_z^s}{|\gamma_e|k_BT}B_z, \label{deltaB}
\end{equation}
where $\gamma_e$ is the gyromagnetic ratio of electrons and $n^s$ is the density of nuclear isotope $s$ on the surface. Given that under the general experimental condition $\vert B_n\vert\ll \vert B_z\vert$ and $|r|\ll 1$ where $r=g_z\mu_B |B_z|/2\sqrt{2\hbar v_F^2 e|B_z|}$, it can be shown that the change of gap at filling factor $\nu$ due to the change of Overhauser field $\delta B_n$ is given by
\begin{equation}
\delta \Delta_L^{\nu}\approx-\frac{g_z \mu_B |\delta B_n|}{2}\times \left\lbrace \begin{array}{ll}
r(\frac{1}{\sqrt{|\nu|+\frac{1}{2}}}-\frac{1}{\sqrt{|\nu|-\frac{1}{2}}}), & |\nu|=n+\frac{1}{2}\\
r+2\nu, & |\nu|=\frac{1}{2}\\

\end{array}\right. \label{LLchange}
\end{equation}
where $n$ is a positive integer. Therefore, by performing NMR, one expects to observe the change of longitudinal conductivity, given by Eqs. (\ref{deltaSigma})(\ref{deltaB}) and (\ref{LLchange}). We point out that in the above analysis we have assumed that the bulk thermal excitation is negligible and ignored the possible effect of the flip-flop processes on the surfaces.

\section{Discussion and conclusion} \label{SecDis}

In this work, we propose to use NMR techniques to both test the topological nature of the surface edge states and control transport in TIs. In 2-D TIs, due to spin-momentum locking of the edge states, there could be significant change in the backscattered current when the nuclei are at resonance.  The change is sensitive to the angle of the polarization axis of the nuclear spins. The phenomenon relies on the flip-flop process between electrons and nuclei as well as current-induced dynamic nuclear polarization on the edge. We show that the change in the current may be two orders of magnitude smaller for non-helical states than for their helical counterparts. As the response is isotropic for the non-helical states unless the hyperfine coupling is anisotropic, this is an alternative approach to characterize the nature of suspicious edge states that may accidentally exist in non-topological systems. For example, it was recently reported that in the topologically trivial phase of InAs/GaSb quantum well, a 2-D TI, there exists an edge current that is believed to be non-helical \citep{Nichele_2016,Sazgari2019}.  For the InAs/GaSb quantum well we estimate that the conductance change percentage for a \mbox{1 $\mu m$} long edge is of order $0.01\%$ when isotope $^{115}$In is depolarized (see Appendix \ref{App4}) so our proposal may be detectable in experiments \citep{Schliemann_2003,Du2015}. 

For a 3-D TI wire, we demonstrate that similar phenomena also exist due to the helicity of 2-D Dirac fermions. Even though the models we use are idealized, especially the 3-D TI Hamiltonian of Eq. (\ref{3DH}), we believe that our prediction that the magnitude of nuclear polarization is sensitive to the magnetic field direction, and that there will be significant change in conductance at the nuclear resonant frequencies, will hold in both clean and mesoscopic systems. Furthermore, we find that RDNMR can be realized in a 3-D TI that is in the quantum Hall regime. This is accomplished by modifying the carrier densities of the Landau levels, through changing the Overhauser field, which then alters the longitudinal conductivity. Using parameters taken from Refs. \onlinecite{Mukhopadhyay2015,Liu2010}, we estimate that when $^{209}$Bi is depolarized in a field $B_z\sim 10\, {\rm T}$ at temperature $T\sim 0.1 \, {\rm K}$, $|\delta B_n|$ is of order \mbox{0.1 T} so the change of Landau gap at filling factor $\nu=\pm 1/2$ is of order 10 $\mu$eV, which is comparable to the temperature and hence should be observable. This idea is similar to RDNMR in a 2-D electron gas in the quantum Hall regime \cite{Desrat2002}, but the Landau level spectrum of the two systems differ significantly. We also emphasize that 2-D Dirac fermions on TI surfaces are different from those of graphene, as in graphene the Zeeman field only results in a relative shift of the energy levels of the different spin species. Therefore the RDNMR experiment in a 3-D TI may reveal the unique spectrum of surface Dirac fermions. 

\begin{acknowledgments}
The authors thank W. Desrat, D. Maude, A. D. Maestro and A. Gromov for useful discussions. This work was supported in part by U.S. National Science Foundation grants OIA-1921199 and OMA-1936221. 
\end{acknowledgments}
\appendix

\section{Derivation of backscattering current and nuclear polarization in the case of 1-D helical/non-helical fermions}\label{App1}
In this appendix, for simplicity we suppress all the superscripts labelling nuclear isotopes. The change rate of nuclear population $P_m^{h/nh}$ is given by Pauli's master equation
\begin{equation}
\frac{d P_m^{h/nh}}{dt}=\sum_{m^\prime} \left( W_{m,m^\prime}^{h/nh}P_{m^\prime}^{h/nh}-W_{m^\prime,m}^{h/nh}P_m^{h/nh}\right), \label{PauliME}
\end{equation}
where $W_{m^\prime,m}^{h/nh}$, the scattering rate of nucleus in the environment of helical/non-helical fermions from state $\vert I,m\rangle$ to state $\vert I,m^\prime\rangle$, is given by Fermi's golden rule:
\begin{widetext}
\begin{equation}
W_{m^\prime,m}^{h/nh}=\frac{2\pi}{\hbar}\sum_{k,\sigma,k^\prime,\sigma^\prime}f_{k\sigma}(1-f_{k^\prime \sigma^\prime})|\langle k^\prime\sigma^\prime,m^\prime\vert H_{hf}\vert k\sigma,m \rangle|^2\delta(\epsilon_{k\sigma}+E_m-\epsilon_{k^\prime \sigma^\prime}-E_{m^\prime}),
\end{equation}
\end{widetext}
where $f_{k\sigma}$ and $\epsilon_{k\sigma}$ are the distribution function/energy of the 1-D fermion in state $\vert k\sigma\rangle$ respectively, and $E_m=-m\Delta$ is the energy of nucleus in state $\vert I,m\rangle$, where $\Delta=\gamma_n \hbar B$ and $\gamma_n$ is the gyromagnetic ratio of specific nuclear isotope. It may be more convenient to rewrite Eq. (\ref{HfH}) in terms of the nuclear spin operators in $z^\prime$ direction $I^\prime_{z,+,-}$, which is related to $I_{z,+,-}$ by $I_i=\sum_j c_{ij}I_{j}^\prime$, where $c_{ij}$ is given by

\begin{equation}
c_{ij}=\left( 
\begin{array}{ccc}
\cos\theta &\frac{\sin\theta}{2}&\frac{\sin\theta}{2}\\
-\sin\theta &\cos^2\frac{\theta}{2} &-\sin^2\frac{\theta}{2}\\
-\sin\theta & -\sin^2\frac{\theta}{2}&\cos^2\frac{\theta}{2}
\end{array}\right)_{ij} 
\end{equation}
in the basis of $z,+,-$.   With the assumption that the distribution of right(left) movers on the edge, given by the left(right) bath's distribution $f_{L(R)}(\epsilon)=(e^{\beta(\epsilon-\mu_{L(R)})}+1)^{-1}$, is unaffected by the weak hyperfine interaction, the scattering rate of nuclei embedded in helical/non-helical Fermi sea reads
\begin{widetext}
\begin{align}
W_{m,m+1}^h=\frac{\pi^2}{h^3v_F^2\beta}(I^-_{m+1})^2 \left[\vert A_{xy} \vert^2\sin^4 \frac{\theta}{2} F(\beta(\Delta-V))+\vert A_{xy} \vert^2\cos^4 \frac{\theta}{2}F(\beta(\Delta+V))+\frac{\vert A_z \vert^2}{2}\sin^2\theta F(\beta\Delta)\right] , \\
W_{m+1,m}^h=\frac{\pi^2}{h^3v_F^2\beta}(I^+_m)^2 \left[\vert A_{xy} \vert^2\sin^4 \frac{\theta}{2}F(-\beta(\Delta-V))+\vert A_{xy} \vert^2\cos^4 \frac{\theta}{2} F(-\beta(\Delta+V))+\frac{\vert A_z \vert^2}{2}\sin^2\theta F(-\beta\Delta)\right],  \label{Wh}
\end{align}
\begin{align}
W_{m,m+1}^{nh}=\frac{\pi^2J^2}{16h\beta}(I^-_{m+1})^2 \left[ F(\beta(\Delta+V))+F(\beta(\Delta-V))+2F(\beta\Delta)\right]\left[\vert A_{xy} \vert^2+(|A_z|^2-|A_{xy}|^2)\frac{\sin^2\theta}{2}\right] , \\
W_{m+1,m}^{nh}=\frac{\pi^2J^2}{16h\beta}(I^+_m)^2 \left[ F(-\beta(\Delta+V))+F(-\beta(\Delta-V))+2F(-\beta\Delta)\right]\left[\vert A_{xy} \vert^2+(|A_z|^2-|A_{xy}|^2)\frac{\sin^2\theta}{2}\right],  \label{Wnh}
\end{align}
\end{widetext} 
where $J$ is non-helical fermion's density of state per length at Fermi levels that counts the spin degeneracy, and we assume it to be roughly a constant over the scale of $V$. We have defined $F(x)=x/(e^x-1)$ and $I^\pm_{m}=\langle I,m\pm 1\vert I_\pm^\prime\vert I,m\rangle=\sqrt{(I\mp m)(I\pm m+1)}$. It is worth mentioning here that the Pauli master equation Eq. (\ref{PauliME}) could also be derived by Lindblad-type quantum master equation if we ignore the interactions between nuclei and makes Born-Markov approximation and secular approximation, which require that the nuclear relaxation time $\tau_{n}\sim[\vert A_{xy} \vert^2/(h^3v_F^2)\times \text{max}(\Delta,V,k_BT)]^{-1}$ is much larger than the electronic relaxation time $\tau_e\sim h/k_BT$ and the intrinsic nuclear time scale $\tau_{\Delta}\sim h/\Delta$  \citep{breuer2002,Probst2015}. We assume that these conditions are satisfied in this paper. At the steady state, the population ratio of nuclei is given by $P_{m+1}/P_m=W_{m+1,m}/W_{m,m+1}$ and therefore
\begin{widetext}
\begin{equation}
\frac{P_{m+1}^h}{P_m^h}=\frac{\sin^4\frac{\theta}{2}F(-\beta(\Delta-V))+\cos^4\frac{\theta}{2}F(-\beta(\Delta+V))+\frac{\vert A_z \vert^2}{2\vert A_{xy} \vert^2}\sin^2\theta F(-\beta\Delta)}{\sin^4\frac{\theta}{2}F(\beta(\Delta-V))+\cos^4\frac{\theta}{2}F(\beta(\Delta+V))+\frac{\vert A_z \vert^2}{2\vert A_{xy} \vert^2}\sin^2\theta F(\beta\Delta)}.
\end{equation}
\begin{equation}
\frac{P_{m+1}^{nh}}{P_m^{nh}}=\frac{F(-\beta(\Delta+V))+F(-\beta(\Delta-V))+2F(-\beta \Delta)}{F(\beta(\Delta-V))+F(\beta(\Delta+V))+2F(\beta\Delta)}
\end{equation}
\end{widetext} 
The backscattering current due to specific isotope is given by the rate of change in the number of left or right movers scattered by that isotope, which is \cite{Lunde2012}
\begin{equation}
I_{bs}=e\sum_{k_+,k_-,\sigma,\sigma^\prime}\left( \Gamma_{k_-\sigma^\prime,k_+\sigma}-\Gamma_{k_+\sigma,k_-\sigma^\prime}\right)\text{sgn}(V),  \label{Ibs}
\end{equation}
where $\Gamma_{b,a}$, the rate for electrons to scatter from state $a$ to $b$ by specific kind of isotope reads
\begin{widetext}
\begin{equation}
\Gamma_{k^\prime\sigma^\prime,k\sigma}=\frac{2\pi}{\hbar}f_{k\sigma}(1-f_{k^\prime \sigma^\prime})N_n\sum_m P_m|\langle k^\prime\sigma^\prime,m^\prime\vert H_{hf}\vert k\sigma,m \rangle|^2\delta(\epsilon_{k\sigma}+E_m-\epsilon_{k^\prime \sigma^\prime}-E_{m^\prime}),
\end{equation}
\end{widetext}
where $H_{hf}$, which is given by Eq. (\ref{HfH}), should be modified to $H_{hf}-A_z\overline{I_z}S_z\delta(r)$ because for spin-conserving process it is the fluctuation of spins scattering particles (see Appendix \ref{App3} for detailed discussion). 
Therefore, the backscattering current on a single edge for helical/non-helical fermion can be easily obtained:
\begin{widetext}
\begin{align}
I_{bs}^h=\alpha_h\frac{e}{h}\frac{\text{sgn}(V)}{\beta}\left[ F(-\beta(\Delta+V))\cos^4\frac{\theta}{2}\times\overline{(I^+_m)^2}+F(\beta(\Delta-V))\sin^4\frac{\theta}{2}\times\overline{(I^-_m)^2}\right. \nonumber\\
-F(\beta(\Delta+V))\cos^4\frac{\theta}{2}\times\overline{(I^-_m)^2}-F(-\beta(\Delta-V))\sin^4\frac{\theta}{2}\times\overline{(I^+_m)^2}\nonumber\\
\left. +F(-\beta V)\sin^2\theta\times\overline{(I^z_m)^2}-F(\beta V)\sin^2\theta\times\overline{(I^z_m)^2}\right],
\end{align}
\begin{align}
I_{bs}^{nh}=\alpha_{nh}\frac{e}{h}\frac{\text{sgn}(V)}{\beta}\left\lbrace  \overline{(I_m^+)^2} \left[  F(-\beta(\Delta+V))-F(-\beta(\Delta-V))\right]  \left[ (1-\frac{\sin^2\theta}{2})+\frac{|A_z|^2}{|A_{xy}|^2}\frac{\sin^2\theta}{2}\right]\right. \nonumber\\ +\overline{(I_m^-)^2} \left[  F(\beta(\Delta-V))-F(\beta(\Delta+V))\right]  \left[ (1-\frac{\sin^2\theta}{2})+\frac{|A_z|^2}{|A_{xy}|^2}\frac{\sin^2\theta}{2}\right]\nonumber\\ \left.+2\overline{(I_m^z)^2} \left[  F(-\beta V)-F(\beta V)\right] \left( \sin^2\theta+\frac{|A_z|^2}{|A_{xy}|^2}\cos^2\theta\right) \right\rbrace,
\end{align}
\end{widetext}
where $\alpha_{nh}=\frac{\pi^2 N_n J^2 |A_{xy}|^2}{16}$ and $I^z_m=m-\overline{m}$.

Notice that we could also include the effective macroscopic magnetic field seen by nuclei due to the current-induced electronic magnetization at the edge, which gives rise to an additional Knight shift term. In the presence of such term, we can always choose a new quantization axis that makes the Hamiltonian of nuclei diagonal and similar analysis holds.

\section{Derivation of backscattering current and nuclear polarization in 3-D TI}\label{App2}
We consider the Hamiltonian given by Eq. (\ref{3DH}) with quantization condition $k_t=(n+1/2)\pi/(L_y+L_z)$. Without losing generality, we assume that the Fermi level lies in the conduction band. In the limit $V\gg k_BT\gg \Delta$, similar to the case in Appendix \ref{App1}, the transition rate of nuclei can be obtained by Fermi's golden rule
\begin{equation}
W^s_{m,m+1}=\frac{2\pi }{\hbar}(A^s)^2V\sum_{n,n^\prime}(I_{m+1}^-)^2|\langle \phi_{n^\prime} |\frac{S^{\hat{n}}_+}{2} | \theta_n \rangle|^2 \frac{J_n}{2}\frac{J_{n^\prime}}{2}, \label{Wm}
\end{equation}
\begin{equation}
W^s_{m+1,m}=\frac{2\pi }{\hbar}(A^s)^2V\sum_{n,n^\prime}(I_m^+)^2|\langle \phi_{n^\prime} | \frac{S^{\hat{n}}_-}{2}| \theta_n \rangle|^2 \frac{J_n}{2}\frac{J_{n^\prime}}{2}, \label{Wp}
\end{equation}
where we have assumed $J_n$ and spinors are almost fixed in the range $E_F-V/2<E<E_F+V/2$. $\phi_{n^\prime}$ is the angle between $\vec{k}$ and $k_x$ axis satisfying $\pi/2<\phi_{n^\prime}<3\pi/2$ (see Fig. \ref{3Dsch}\hyperlink{3Dsch}{(c)}) and $\vert \phi_{n^\prime}\rangle$ is the corresponding spinor. $S^{\hat{n}}_{+(-)}$ denotes the raising (lowering) operator in the general $\hat{n}$ direction. For simplicity, from here we only consider the case that external magnetic field is along the principal axes and we focus on the top surface shown in Fig. \ref{3Dsch}\hyperlink{3Dsch}{(a)}. The matrix element of various spin operators along different principal axes can be calculated to be
\begin{align}
 |\langle \phi_{n^\prime} | S^y_-| \theta_n \rangle|^2&=\sin^2\frac{\theta_n}{2}\cos^2\frac{\phi_{n^\prime}}{2},\\
 |\langle \phi_{n^\prime} | S^y_+| \theta_n \rangle|^2&=\cos^2\frac{\theta_n}{2}\sin^2\frac{\phi_{n^\prime}}{2},\\
 |\langle \phi_{n^\prime} | S^y| \theta_n \rangle|^2&=\frac{1}{4}\cos^2(\frac{\phi_{n^\prime}+\theta_n}{2}),\\
  |\langle \phi_{n^\prime} | S^z_-| \theta_n \rangle|^2&=\frac{1}{4},\\
  |\langle \phi_{n^\prime} | S^z_+| \theta_n \rangle|^2&=\frac{1}{4},\\
  |\langle \phi_{n^\prime} | S^x_-| \theta_n \rangle|^2&=\sin^2(\frac{\theta_n}{2}+\frac{\pi}{4})\cos^2(\frac{\phi_{n^\prime}}{2}+\frac{\pi}{4}),\\
  |\langle \phi_{n^\prime} | S^x_+| \theta_n \rangle|^2&=\cos^2(\frac{\theta_n}{2}+\frac{\pi}{4})\sin^2(\frac{\phi_{n^\prime}}{2}+\frac{\pi}{4}).
\end{align}
 $S^x_{-(+)}$ denotes the lowering (raising) operator in the $x$ direction and same holds for other operators; especially, $S^y\equiv \sigma_y/2$. Combining the above equations one can show that
\begin{equation}
    \frac{P_{m+1}}{P_m}=\frac{W_{m+1,m}}{W_{m,m+1}}=
    \left\lbrace 
    \begin{array}{ll}
    1,&  \hat{n}=\hat{x} \text{ or } \hat{z}\\
    (\frac{\gamma}{1-\gamma})^2,& \hat{n}=\hat{y}
    \end{array}
    \right..
\end{equation}

Similar to Eq. (\ref{Ibs}), given positive $V$, the backscattering current in 3-D TI is given by
\begin{equation}
    I_{bs}=e\sum_{k_+,k_-,n,n^\prime} (\Gamma_{k_- n^\prime,k_+ n}-\Gamma_{k_+ n,k_-n^\prime}).
\end{equation}
In the limit $\Delta\ll k_BT\ll V$ the last term vanishes and above equation becomes
\begin{equation}
I_{bs}=eN_n\sum_m P_m(W_{m+1,m}+W_{m-1,m}+W_{m,m})
\end{equation}
where
\begin{equation}
    W_{m,m}=\frac{2\pi }{\hbar}(A^s)^2V\sum_{n,n^\prime}(I_m^z)^2|\langle \phi_{n^\prime} |S^{\hat{n}}_z | \theta_n \rangle|^2 \frac{J_n}{2}\frac{J_{n^\prime}}{2}, \label{Wz}
\end{equation}
counts the rate of scattering events conserving nuclear spins. Combining the above equations one can obtain Eq. (\ref{deltaG3D}).
\section{Justification of Fermi's golden rule with non-equilibrium Green's function}\label{App3}
It is known that in the semi-classical regime, the conductivity in disordered system to the lowest order can be calculated with Fermi's golden rule. However, for a system with highly dense magnetic scatters, some attention should be paid when using Fermi's golden rule to calculate the conductivity. For clarification, in this appendix we sketch the verification of Fermi's golden rule used in Appendix \ref{App1} with non-equilibrium Green's function (NEGF). Specifically, we demonstrate why the hyperfine interaction $H_{hf}$ should be replaced by $H_{hf}^\prime\approx \sum_a A(I^a-\overline{I^a})\cdot S^a$ when calculating the matrix elements. Consider a translationally invariant system with Hamiltonian $H_0$ interacting with nuclei that have arbitrary spin temperature through the isotropic hyperfine interaction $V$ is $V=\sum_i(\tilde{A}_i^{s_i})_{k,k^\prime}\vec{I}^{s_i}\cdot\vec{S}_{\alpha\beta}c_{k\alpha}^\dagger c_{k^\prime\beta} $, where $i$ labels the site of nuclei, $s_i$ labels the corresponding isotopic species, $k,k^\prime$ is the momentum, $\alpha,\beta$ is the spin index, $(\tilde{A}_i^{s_i})_{k,k^\prime}=\frac{A^{s_i}e^{i(k^\prime-k)R_i}}{L^D}$, $D$ is the dimensionality of the system and $L$ is the length of each dimension. In the Keldysh formalism, the Green's function reads
\begin{widetext}
\begin{multline}
     G_{k\alpha;k^\prime\alpha^\prime}=\delta_{k,k^\prime}(G_0)_{k\alpha;k\alpha^\prime}+\sum_{i}(G_0)_{k\alpha;k\alpha_1}(\tilde{A}_i^{s_i})_{k,k^\prime}\overline{(I^{s_i})^a}{S}^a_{\alpha_1\alpha_2}(G_0)_{k^\prime\alpha_2;k^\prime\alpha^\prime}\\+ \sum_{i,j}(G_0)_{k\alpha;k\alpha_1}(\tilde{A}_i^{s_i})_{k,k_1}S^a_{\alpha_1\alpha_2}(G_0)_{k_1\alpha_2;k_1\alpha_3}(\tilde{A}_j^{s_j})_{k_1,k^\prime}S^b_{\alpha_3\alpha_4}(G_0)_{k^\prime\alpha_4;k^\prime\alpha^\prime}\overline{\mathcal{T}_c(I^{s_i})^a(I^{s_j})^b}+...   ,    \label{GFexpansion}
\end{multline}
\end{widetext}
where $G_0$ is the bare Green's function, $a,b$ labels the component of spin, $\mathcal{T}_c$ is the time-ordering operator along the Keldysh contour and $\overline{(...)}$ denotes the thermal average for nuclei. For convenience we have suppressed all the time arguments and their integral; we note that, for example, in general spin operators in terms such as $\overline{\mathcal{T}_c(I^{s_i})^a(I^{s_j})^b}$ should be evaluated at different time on the Keldysh contour. The summation symbol for repetitive indices is also implicitly indicated from now on. To represent Eq. (\ref{GFexpansion}) in the form of Dyson's equation, one could first decompose all the expectation values to the corresponding cumulants, e.g. $\overline{ \mathcal{T}_c(I^{s_i})^a(I^{s_j})^b}=\overline{\mathcal{T}_c \delta(I^{s_i})^a\delta(I^{s_j})^b}+\overline{ (I^{s_i})^a}\cdot\overline{(I^{s_j})^b}$, where $\delta(I^{s_i})^a=(I^{s_i})^a-\overline{ (I^{s_i})^a}$. Next, we ignore the correlation between nuclear spins at different sites, therefore terms such as $\overline{\mathcal{T}_c \delta(I^{s_i})^a\delta(I^{s_j})^b}$ vanish unless $i=j$. Lastly, since we are less interested in the physics with specific configuration of nuclear isotopes, we take ensemble average of Eq. (\ref{GFexpansion}) with respect to different realization of nuclear isotopes. After these procedures, Eq. (\ref{GFexpansion}) can be written in the form of Dyson series
\begin{equation}
    G_{k\alpha;k \alpha^\prime}=(G_0)_{k\alpha;k \alpha^\prime}+(G_0)_{k\alpha;k \alpha_1} \Sigma_{k\alpha_1;k\alpha_2} G_{k\alpha_2;k \alpha^\prime}, \label{Dyson}
\end{equation}
where the first-order and second-order self-energy are respectively
\begin{equation}
    (\Sigma^1)_{k\alpha;k\beta}=\sum_{s}n^s A^s\overline{(I^s)^a}(S^a)_{\alpha\beta},
\end{equation}
\begin{multline}
    (\Sigma^2)_{k\alpha;k\beta}=\sum_{s}\frac{n^s (A^s)^2 \overline{\mathcal{T}_c\delta(I^s)^a \delta(I^s)^b}}{L^D}\\\times(S^a)_{\alpha\gamma}(G_0)_{k^\prime\gamma;k^\prime\delta}(S^b)_{\delta\beta},    
\end{multline}
in which $n^s$ is the density of specific isotope. Notice that the momentum of Green's function is conserved in Eq. (\ref{Dyson}) after performing ensemble average. The first-order self-energy gives rise to the slight modification of spectrum due to nuclear field (Overhauser shift) which is neglected here, while the imaginary part of the second-order self-energy gives the decay rate due to the hyperfine interaction, which can be shown to be equivalent to Fermi's golden rule if one relates it to the retarded self-energy in real-time domain. Note that it is $\overline{\delta(I^s)^a \delta(I^s)^b}$ rather than $\overline{(I^s)^a (I^s)^b}$ entering the second-order self-energy, which justifies that in Appendix \ref{App1} when using Fermi's golden rule hyperfine interaction should be replaced by an effective interaction $H_{hf}^\prime\approx \sum_a A(I^a-\overline{I^a})\cdot S^a$. 

\section{Estimate of magnitude of RDNMR signal} \label{App4}
In this appendix, we give an estimate of the conductance change in the InAs/GaSb quantum well as an example. We assume the hyperfine coupling $A_{1D}$ is isotropic for simplicity. In reality, the localized edge state not only spreads in the quantum well but also extends into the bulk. Therefore the edge state has a finite cross section $S_{\text{edge}}$ and interacts with all the nuclei overlapping with the wavefunction. The effective 1-D hyperfine coupling constant is given by
\begin{equation}
    A_{1D}=\frac{A_{3D}}{S_{\text{edge}}}
\end{equation}
where $A_{3D}$ is the 3-D hyperfine coupling constant estimated by its bulk value. The number of nuclei of isotope $^{115}$In that backscatter Dirac fermions is given by
\begin{equation}
    N_n^s=\alpha \frac{L_xS_{\text{edge}}}{V_c}
\end{equation}
where $V_c$ is the unit cell volume of InAs. The coefficient $0<\alpha<1$ takes into account that Indium atoms do not exist in GaSb and $^{115}$In has a natural abundance of $\sim 95.7\%$. The conductance change in Eq. (\ref{2Dchange}) can thus be written as
\begin{equation}
    \frac{\delta G}{G}=\alpha\frac{2\pi^2 (A_{3D})^2L_x}{3h^2v_F^2V_cS_{\text{edge}}}I(I+1).
\end{equation}
We obtain the estimated change of conductance of $0.01 \%$ quoted in the main text by setting $I=9/2$, $S_{\text{edge}}\approx 150\, \text{nm}^2$, $V_c\approx 55.6\,  \text{\AA}^3$, $A_{3D}\approx 3780 \, {\rm \mu eV}\cdot\text{\AA}^3$ \cite{Schliemann_2003}, $v_F\approx 1.5\times 10^4 \, {\rm m/s}$ \cite{Du2015} and setting $\alpha=1$.

\newpage
\bibliographystyle{apsrev4-2}
\bibliography{RDNMRRef}

\end{document}